\pdfoutput=1
\documentclass[cits]{JINST}
\usepackage{graphicx}
\usepackage{dcolumn}
\usepackage{bm}
\usepackage{verbatim}
\usepackage{units}
\usepackage{subfigure}
\usepackage{amsmath}
\usepackage{lineno}
\title{An angular-selective electron source for the KATRIN experiment}

\author{M. Beck$^{a, b}$\thanks{Corresponding author, present address: Helmholz-Institut Mainz, Germany}, K. Bokeloh$^a$\thanks{n\'ee K. Hugenberg}, H. Hein$^a$, S. Bauer$^a$\thanks{Physikalisch-Technische Bundesanstalt, Bundesallee 100, D-38116 Braunschweig, Germany}, H. Baumeister$^a$, J. Bonn$^b$\thanks{Deceased}, H.-W. Ortjohann$^a$, B. Ostrick{$^{a, b}$}, S. Rosendahl$^a$, S. Streubel$^a$\thanks{Present address: Max-Planck-Institut f\"ur Kernphysik, D-69029 Heidelberg, Germany}, K. Valerius$^a$\thanks{Present address: KCETA, Karlsruhe Institute of Technology, P.O. Box 3640, D-76021 Karlsruhe, Germany}, M. Zbo\v ril$^{a, c}$\thanks{Present address: Physikalisch-Technische Bundesanstalt, Bundesallee 100, D-38116 Braunschweig, Germany}~and C. Weinheimer$^a$\\
\llap{$^a$}Institut f\"ur Kernphysik, Westf\"alische Wilhelms-Universit\"at,\\
 D-48149 M\"unster, Germany\\
\llap{$^b$}Institut f\"ur Physik, Johannes Gutenberg-Universit\"at,\\
 D-55099 Mainz, Germany\\
\llap{$^c$}Nuclear Physics Institute ASCR,\\
 CZ-25068 \v{R}e\v{z} near Prague, Czech Republic\\
E-mail: \email{beck005@uni-mainz.de}}

\abstract{
The KATRIN experiment is going to search for the average mass of the electron antineutrino with a sensitivity of $\unit[0.2]{eV/c^2}$. It uses a retardation spectrometer of MAC-E filter type to accurately measure the shape of the electron spectrum at the endpoint of tritium beta decay. In order to achieve the planned sensitivity the transmission properties of the spectrometer have to be understood with high precision for all initial conditions. For this purpose an electron source has been developed that emits single electrons at adjustable total energy and adjustable emission angle. The emission is pointlike and can be moved across the full flux tube that is imaged onto the detector. Here, we demonstrate that this novel type of electron source can be used to investigate the transmission properties of a MAC-E filter in detail.}


\begin{document}

\newcommand{\KE}{KATRIN experiment}
\newcommand{\eg}{{\it e.g.}}
\newcommand{\ie}{{\it i.e.}}
\newcommand{\etal}{{\it et al.}}

\section{Introduction}
\label{sec:intro}

The KArlsruhe TRItium Neutrino experiment (KATRIN, \cite{kdr,bec10}) will search for the mass of the electron antineutrino with a sensitivity of $0.2~\mathrm{eV/c^2}$. It will probe most of the cosmologically relevant region of degenerate neutrino masses \cite{{hannestad14},dre13,lesgourgues-pastor-review2,otten-weinh-review,lesgourgues-pastor-review1}. In order to achieve this goal the endpoint region of the $\beta$-decay spectrum of tritium will be investigated with KATRIN. The shape of the spectrum in this region depends on the neutrino mass. This shape will be measured precisely using an electrostatic spectrometer of MAC-E filter type (electrostatic filter with magnetic adiabatic collimation \cite{picard-nimb,lobashev85}), which at the same time has a high solid angle acceptance (up to $2\pi$) and an excellent energy resolution. This method was already used at the Mainz and Troitzk neutrino mass experiments, which have set the until now best upper laboratory limits of $\sim2~\mathrm{eV/c^2}$ on the mass of the electron antineutrino \cite{kra05,lob99}.

\subsection{Measurement principle}
\label{sec:mace}

\begin{figure*}[htb]
 \centering
 \includegraphics[width=\textwidth]{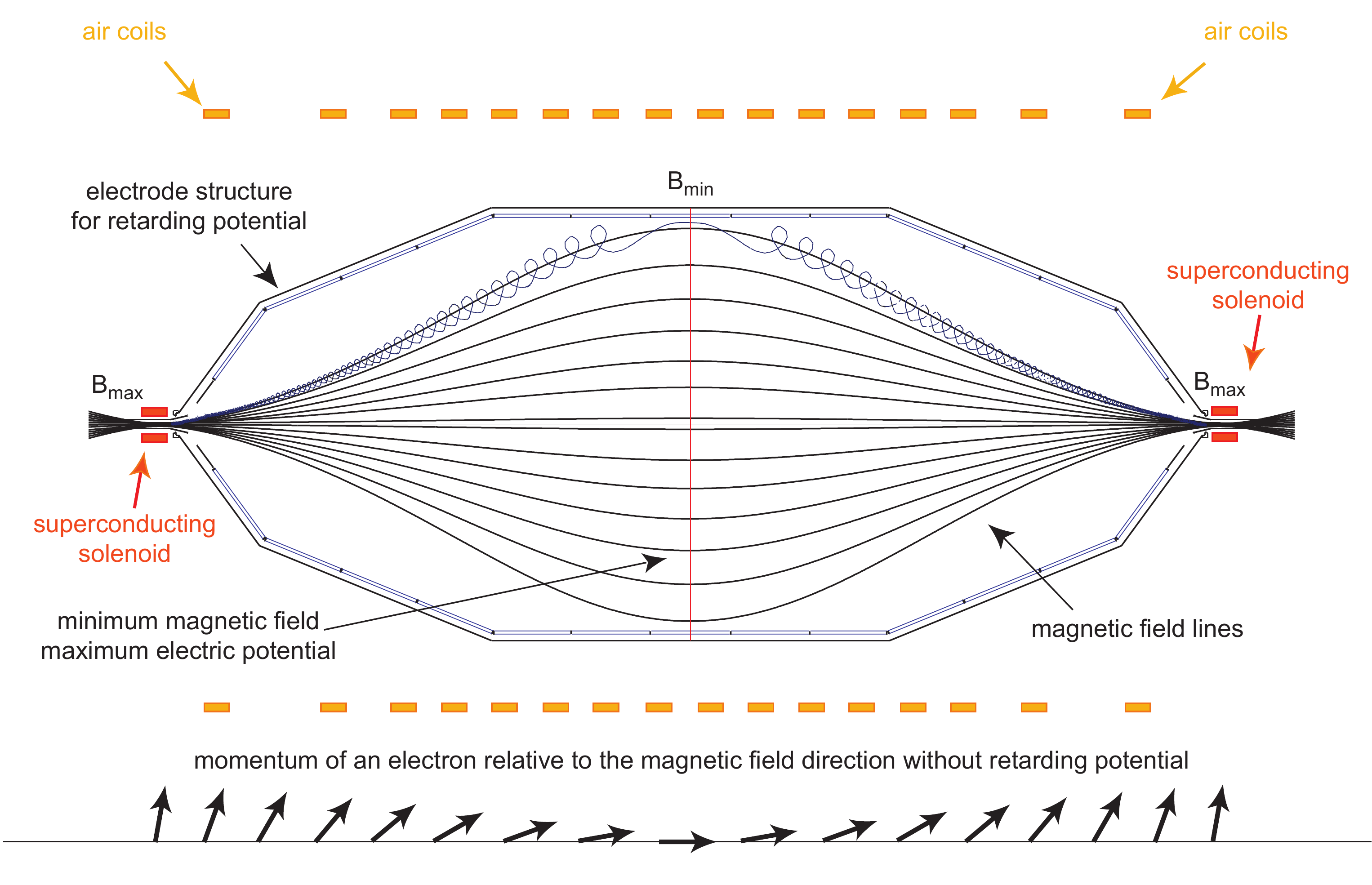}
 \caption[Schematic drawing of a MAC-E filter]{Schematic drawing of the KATRIN main spectrometer as an example of a MAC-E filter. An exaggerated cyclotron motion of an electron is shown (spiraling line). The arrows at the bottom of the figure give the relative momentum of an example electron. The spectrometer has a length of $23~\mathrm{m}$ and a diameter of nearly $10~\mathrm{m}$.}
 \label{fig:MACE}
\end{figure*}

\begin{figure}[t]
 \centering
 \includegraphics[width=0.6\textwidth]{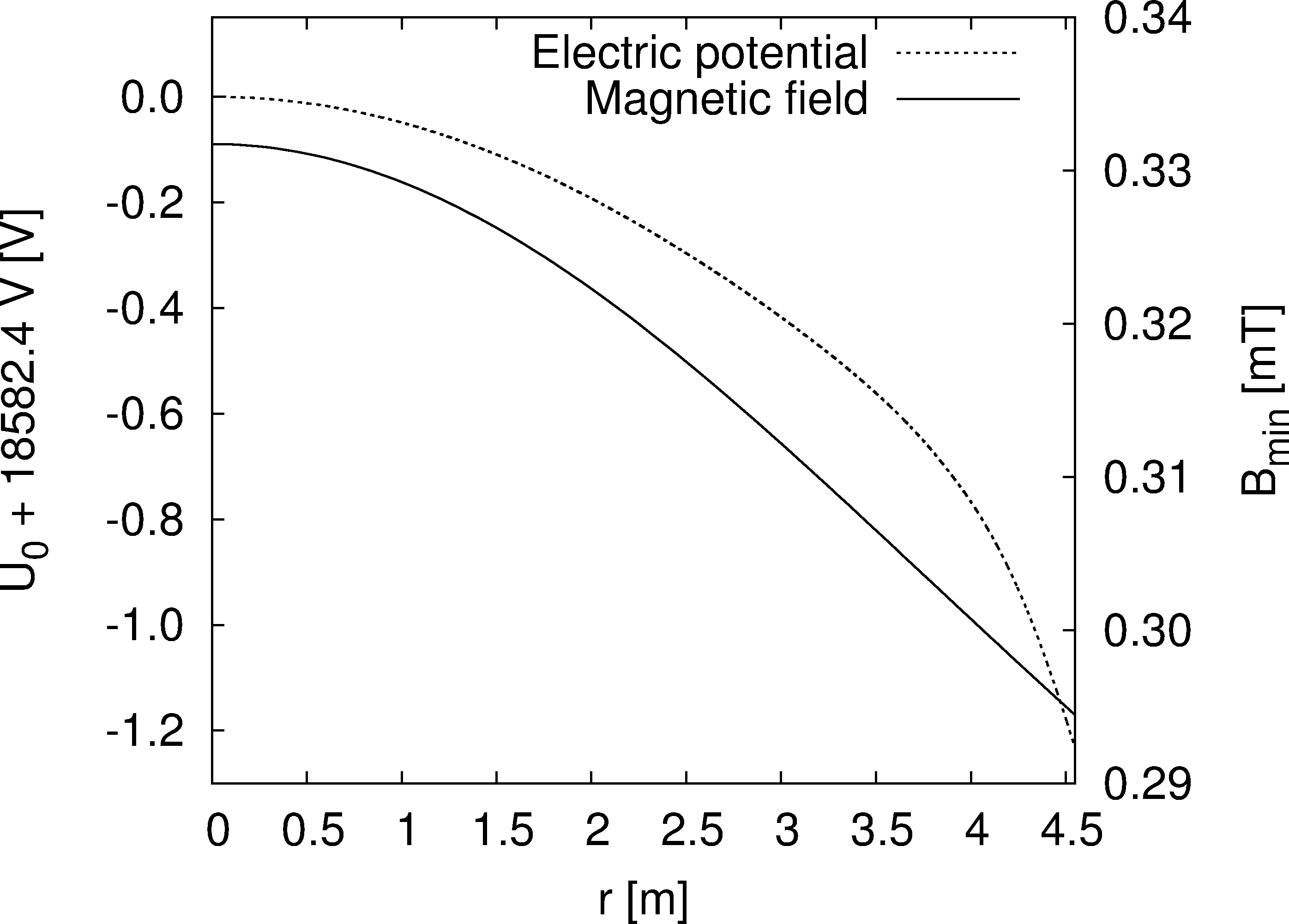}
 \caption[Radial dependence of E- and B-field]{Dependence of the electric potential and magnetic field strength in the analyzing plane on the radial distance from the spectrometer axis $r$. (Figure adapted from \cite{diss_bokeloh}.)}
 \label{fig:ebfields}
\end{figure}

In order to reach the sensitivity goal of KATRIN the systematic uncertainties have to be investigated in detail. Besides the precise knowledge of the properties of the source -- at KATRIN a windowless gaseous tritium source -- the understanding of the transmission properties of the KATRIN main spectrometer is of eminent importance. A spectrometer of MAC-E filter type consists of an electrostatic retardation filter combined with an inhomogeneous magnetic guiding field.
 Figure~\ref{fig:MACE} shows the KATRIN main spectrometer schematically, with the two superconducting solenoids at its ends, which produce a strong magnetic field $B_\mathrm{max}$ that decreases towards the center of the spectrometer by about four orders of magnitude to a minimum value $B_\mathrm{min}$. An air coil system is used to fine tune the magnetic field. The vacuum tubes inside the two main solenoids are on ground potential whereas the vessel of the spectrometer is on a high negative potential. Thus, the absolute value of the electric potential increases from ground potential inside the high magnetic field of the first solenoid towards the center of the spectrometer where it reaches a maximum and decreases again towards the second solenoid at the exit of the spectrometer. The electric potential inside the spectrometer is fine tuned using an inner electrode system consisting of a double layer of wires, which are slightly more negative than the vessel of the spectrometer. The retardation potential $U_0$ is defined by the potential of the innermost wire layer. Due to the finite dimensions of the spectrometer the effective retardation potential shows a tiny radial dependence (Fig.~\ref{fig:ebfields}).

The decrease of the magnetic field strength from the spectrometer entrance towards its center results in an expansion of the magnetic flux tube, visualized in Fig.~\ref{fig:MACE} by a widening of the magnetic field lines. Electrons starting from a source in the high magnetic field will in general have both a momentum component parallel and a component perpendicular to the magnetic field lines. This latter, transversal component of their motion forces the electrons into a cyclotron motion around the field lines whereas the parallel component will transport the electrons emitted in forward direction along the magnetic field lines into the spectrometer. The electric and magnetic fields are shaped such that the electrons move adiabatically for the electron energies present at KATRIN, meaning that the relative change of the magnetic field strength is small during the cyclotron motion, \ie, $| \frac{1}{B}\frac{\mathrm{d}B}{\mathrm{d}z} | \ll \frac{\omega_\mathrm{c}}{v_\parallel}$ with $\omega_\mathrm{c}$ the cyclotron frequency and $v_\parallel$ the axial velocity. Disregarding in this first consideration the retardation potential, both the total kinetic energy and the orbital magnetic moment $\mu$ of the electrons are conserved (see, \eg, \cite{otten-weinh-review,kruit-read}). Thus, with decreasing magnetic field strength the energy stored in the cyclotron motion of the electron, $E_\perp$, will be transformed into axial kinetic energy, $E_\parallel$, according to the relation (see, \eg, ref.~\cite{art_fegun})
\begin{equation} 
 \mu =  \frac{E_\perp}{B} = \mathrm{const.}
\label{equ:magn-moment}
\end{equation}
The adiabatic invariant is considered here in the non-relativistic limit. The relation can be rewritten as
\begin{equation}
\frac{\sin^2\theta_f}{\sin^2\theta_i} = \frac{B_f}{B_i},
\label{equ:angle-transform}
\end{equation} 
where the indices $i$ and $f$ denote initial and final values. The angle $\theta$ describes the direction of the momentum $\vec{p}$ of the electron with respect to the magnetic field $\vec{B}$:
\begin{equation}
\cos \theta = \frac{\vec{p} \cdot \vec{B}}{|\vec{p}| \, |\vec{B}|}. 
\label{equ:theta}
\end{equation}

\begin{figure}[t]
\centering
\includegraphics[width=0.6\textwidth]{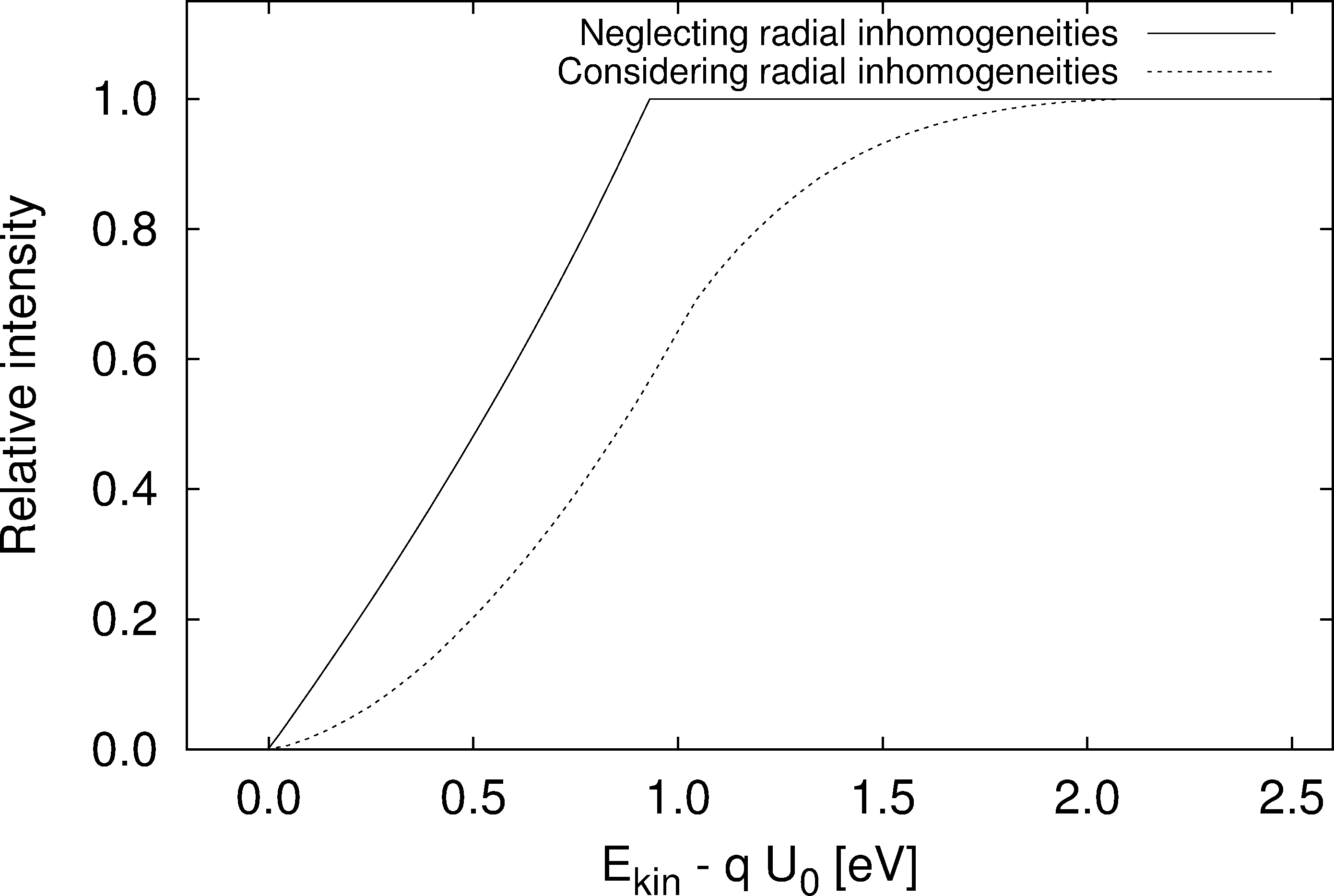}
\caption[Transmission function]{Calculated transmission functions of a MAC-E filter with the KATRIN main spectrometer settings, for an isotropically emitting source of $18.6~\mathrm{keV}$ electrons ({\it cf.} footnote $^1$). Solid line:  idealized configuration, no radial inhomogeneity of retardation potential or magnetic field considered. Dashed line: including radial inhomogeneity of the retardation potential and of the magnetic field with an amplitude (cf. Fig.~\ref{fig:ebfields}) of approximately $\unit[1.2]{V}$ and $\unit[4\cdot 10^{-5}]{T}$. Due to these field inhomogeneities the width of the transmission function will be increased from the nominal value of $\Delta E = \unit[0.93]{eV}$ to about $\unit[2.4]{eV}$. (Figure adapted from \cite{diss_bokeloh}.) \label{fig:tfunc}}
\end{figure}

\subsection{Resolution of the MAC-E filter}
\label{resolution}

Only the axial energy component $E_\parallel$ can be analyzed with the axial electrostatic retardation field. Therefore, in order to minimize the non-analyzable fraction of the kinetic energy, {\it i.e.} its transversal component, the magnetic field reaches its minimal value $B_\mathrm{min}$ in the same region where the retardation potential $U_0$ reaches its maximum. This ensures that the maximal amount of the transversal energy has been converted into axial energy for the energy analysis. ``Transversal'' and ``axial'' components are defined with respect to the direction of the magnetic field. In the center of the spectrometer $\vec{B}$ and $\vec{E}$ are in good approximation aligned with the spectrometer axis. The region of maximal retardation potential is called ``analyzing plane'' ($z = 0$, vertical line in the center in Fig.~\ref{fig:MACE}). Only those electrons with a sufficient amount of axial kinetic energy, $E_\parallel > qU_0$, with $q=-e$ the charge of the electron, will be able to pass the analyzing plane and get re-accelerated towards the detector at the exit of the spectrometer. Using Eqs.~\ref{equ:magn-moment} and \ref{equ:angle-transform} the condition for transmission can be written as

\begin{equation}
qU_0 < E_\mathrm{kin} \left( 1 - \frac{B_\mathrm{min}}{B_\mathrm{max}}  \sin^2 \theta_m \right),
\label{equ:tcond}
\end{equation}

\noindent with $\theta_m$ being the angle between the momentum of the electron and the magnetic field in the first solenoid at the entrance of the spectrometer. In this simple case this is at the location of maximum magnetic field strength $B_\mathrm{max}$.

The energy resolution $\Delta E$ of the MAC-E filter thus follows directly from Eq.~\ref{equ:tcond} by assuming that an electron starts at $B_\mathrm{max}$ with an initial kinetic energy $E_\mathrm{kin, start}$ that resides entirely in the cyclotron component, {\it i.e.} $\theta_m = 90^\circ$. In this case, $\Delta E$ corresponds to the maximal amount of transverse energy $(E_\perp)_\mathrm{max}$ that is left at the electrostatic analyzing plane of the filter after the adiabatic transformation according to Eq.~\ref{equ:magn-moment}, and which cannot be probed by the analyzing potential:
\begin{equation}
 \Delta E = (E_\perp)_\mathrm{max} = E_\mathrm{kin, start} \cdot \frac{B_\mathrm{min}}{B_\mathrm{max}}.
\label{equ:energyres}
\end{equation}

Consequently, the shape of the transmission function on axis is only determined by the ratio of the maximal and minimal magnetic fields. In the case of an isotropically emitting source in the entrance solenoid of the spectrometer at the maximum magnetic field $B_\mathrm{max}$ the transmission function becomes:

\begin{equation}
T(E_\mathrm{kin, start}, qU_0) = \left\lbrace 
\begin{array}{lll}
0												&\text{for}	& E_\mathrm{kin, start} < qU_0  \\[3mm]
1 - \sqrt{1 - \dfrac{(E_\mathrm{kin, start} - qU_0)}{E_\mathrm{kin, start}}\dfrac{B_\mathrm{max}}{B_\mathrm{min}}} 	& \text{for}	& E_\mathrm{kin, start} \left(1 - \dfrac{B_\mathrm{min}}{B_\mathrm{max}}\right)\leq qU_0 \leq E_\mathrm{kin, start}\\[5mm]
1												& \text{for}	& qU_0 \leq E_\mathrm{kin, start} \left(1 - \dfrac{B_\mathrm{max}}{B_\mathrm{min}}\right).
\end{array}\right.
 \label{equ:transm}
\end{equation}


$\Delta E$ in Eq.~\ref{equ:energyres} therefore also corresponds to the width of the transmission function at a given initial kinetic energy $E_\mathrm{kin, start}$. The resulting transmission function for the default settings\footnote{Note that the maximum magnetic field strength in the overall KATRIN set-up is not reached within the source, with $B_\mathrm{source}=3.6~\mathrm{T}$, or inside the entrance solenoid of the spectrometer, but is instead defined by a so-called ``pinch'' magnet with $B_\mathrm{pinch}=6~\mathrm{T}$ located at the detector side of the spectrometer. This results in a maximal accepted angle $\theta_i = \arcsin\sqrt{\frac{B_\mathrm{source}}{B_\mathrm{pinch}}} = 50.8^\circ$.} of KATRIN is shown in Fig.~\ref{fig:tfunc} (solid line) for electron energies of $18.6~\mathrm{keV}$.

The electric and magnetic fields can be calculated numerically based on the configuration of the spectrometer electrodes and magnet coils \cite{glu11a,glu11b}. In the ideal case, the magnetic field strength decreases monotonously when approaching the analyzing plane and rises monotonously when receding from it. Likewise, the electric potential is highest in the analyzing plane and falls off towards both sides. In addition to this pronounced axial inhomogeneity, both the electric retardation potential and the magnetic field strength also exhibit some radial dependence along the analyzing plane (Fig.~\ref{fig:ebfields}). This radial inhomogeneity causes a smearing and broadening of the transmission function (Fig.~\ref{fig:tfunc}, dotted line). To compensate for this broadening KATRIN utilizes a segmented detector which allows to separately investigate electrons that pass the spectrometer at different radial positions.

\begin{figure}[htb]
 \centering
 \includegraphics[width=0.6\textwidth]{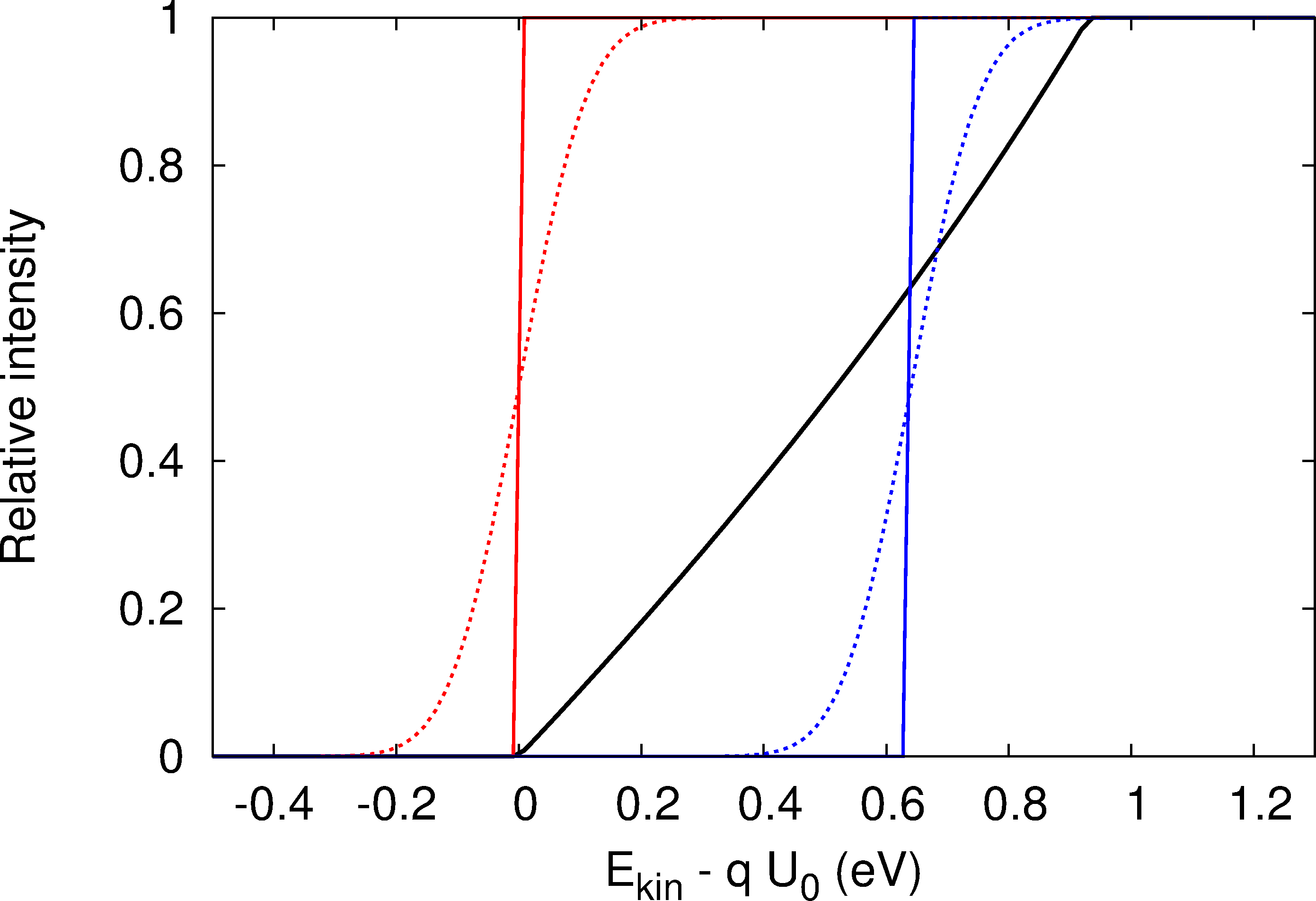}
 \caption[]{Calculated transmission functions for KATRIN. For an isotropic source (black solid line) the transmission function spans the width given by Eq.~\ref{equ:energyres}, {\it i.e.} $0.93~\mathrm{eV}$ (see footnote $^1$). The step functions (red and blue solid lines) represent transmission functions for electrons of sharply defined emission angle ($\theta_i = 0^\circ$ and $40^\circ$). A spread of the initial energy $E_\mathrm{kin, start}$, noise on the electric potential or deviations from the ideal field configuration lead to a broadening of the step function (dotted lines: Gaussian broadening of $\sigma = 0.1~\mathrm{eV}$). (Figure adapted from \cite{diss_bokeloh}.)}
 \label{fig:tfuncsigma}
\end{figure}

Additional broadening of the transmission function, changes to the theoretically expected shape or even transmission losses may arise due to deviations from the ideal fields \cite{art_fegun,diss_kathrin}. These may be caused, \eg, by an imperfect realization of the electrodes or coil geometries that create the electric and magnetic field, or external noise. An example of the effect of the latter is shown in Fig.~\ref{fig:tfuncsigma} for electrons of fixed initial angle $\theta_i$ (dotted lines).

\begin{figure}[htb]
 \centering
 \includegraphics[width=0.7\textwidth]{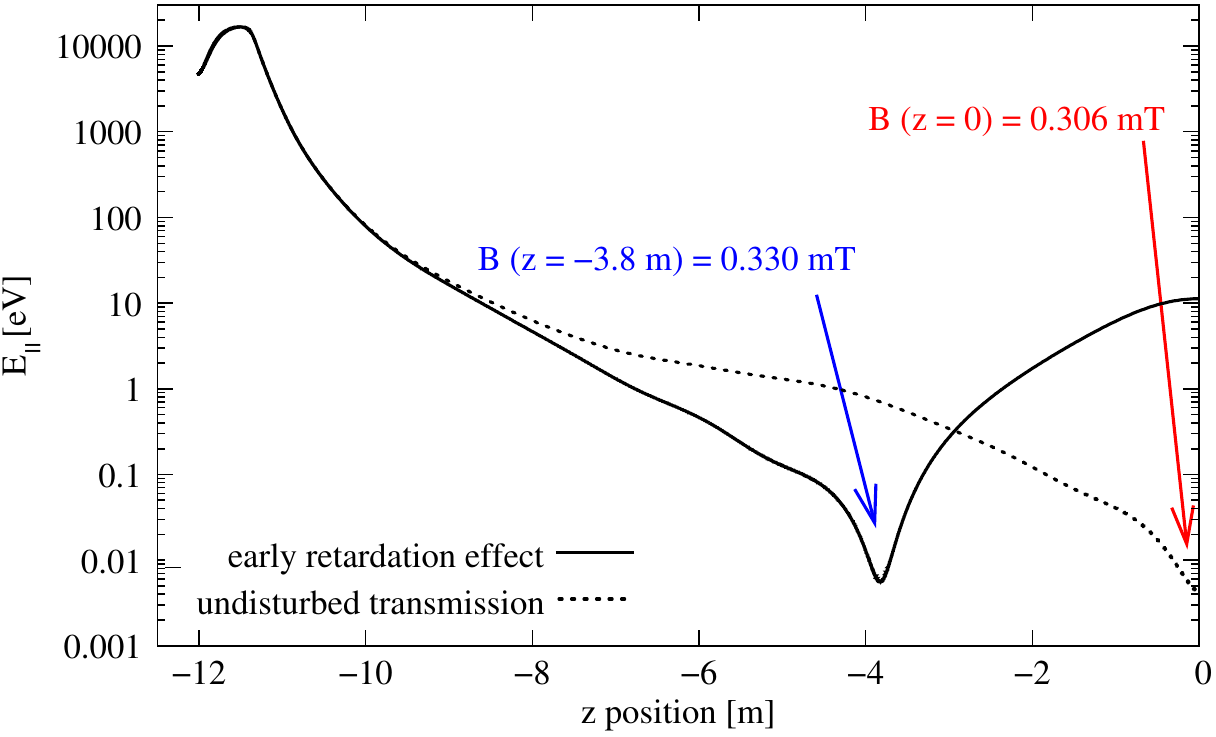}
 \caption[]{Simulation of the effect of early retardation on the axial energy component $E_\parallel$ for an electron at the transmission threshold (solid line). In the ideal case (dotted line), $E_\parallel$ reaches its minimum at $z=0$. (Figure adapted from \cite{diss_kathrin}.)}
 \label{fig:earlyret}
\end{figure}

\begin{figure}[htb]
 \centering
 \includegraphics[width=0.7\textwidth]{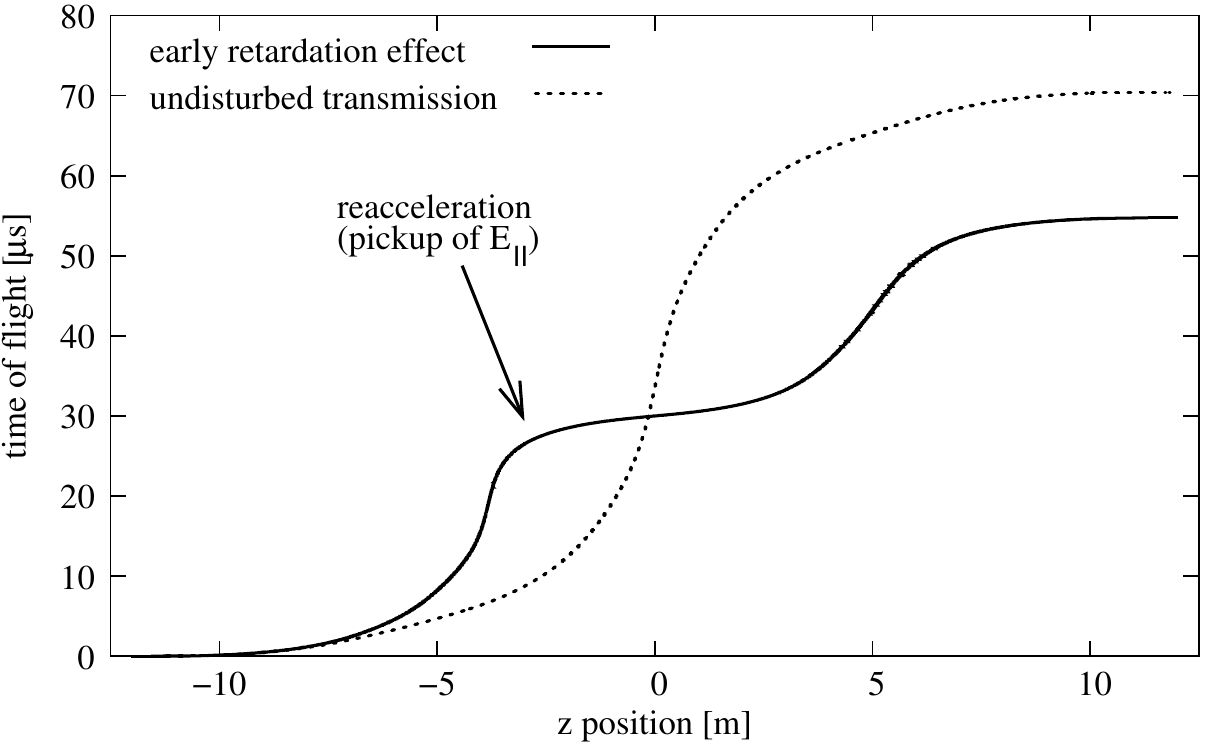}
 \caption[]{Effect of early retardation on the time of flight through the main spectrometer for the simulation in Fig.~\ref{fig:earlyret}. (Figure adapted from \cite{diss_kathrin}.)}
 \label{fig:earlyrettof}
\end{figure}

Deviations from the ideal fields can, however, have a more dramatic effect. They might cause the retardation potential to reach its maximum before the ideal analyzing plane position (Fig.~\ref{fig:earlyret}, solid line). Here the magnetic field strength would still be higher than in the analyzing plane, resulting in an incomplete conversion of radial into axial energy. Electrons with large transveral energy and total energy slightly above the analyzing potential would then be reflected, in contrast to what would be expected according to Eqs.~\ref{equ:tcond} and \ref{equ:transm}. Such an unaccounted loss of electrons would lead to additional systematic uncertainties on the neutrino mass. It is obvious that any such effect will be most pronounced for electrons that move close to the electrodes and that have large initial emission angles $\theta_i$. In a measurement of the transmission function this effect is difficult to observe, since it will just lead to an unspecific broadening of the transmission function due to the incomplete conversion of radial into axial energy at the place of maximum retardation. However, it will show up in a measurement of the time of flight of the electrons (Fig.~\ref{fig:earlyrettof}, and cf. \cite{diss_kathrin}). Consequently, a time-of-flight measurement with a suitable electron source can be used to investigate this effect.

\subsection{Requirements for a calibration source}
\label{requirements}

Using a suitable electron source connected directly to the spectrometer deviations of the transmission function from its ideal shape can be probed and their origins can be identified.
\begin{itemize}
\item In order to unfold some of the effects like the field inhomogeneity in the analyzing plane and field deviations due to imperfections in the electrode structure or the air coil system, a quasi-pointlike electron source is needed, which has to be movable across the magnetic flux tube that leads to the detector. Such an electron source allows to check the fields separately at all radii and for all detector segments. This is most important for field lines passing close to the electrodes, where field deviations will have the largest effect.
\item In order to test the adiabaticity of the motion at various excess energies $E_\mathrm{kin,\, start} - qU_0$ the electron energy has to be variable.
\item Electrons with well-defined, selectable transversal energy are needed to sensitively test the conversion of transversal into axial energy. This will provide a much higher sensitivity to deviations of the electric field than using electrons from a source that emits electrons with a wide spread of transversal energies (see Fig.~\ref{fig:tfuncsigma}).
\item Finally, a pulsed electron source that emits single electrons will allow time-of-flight studies of the spectrometer which not only probe the fields in the analyzing plane but integrate over all fields along the trajectory of the electron:

\begin{equation}
t_\mathrm{stop} - t_\mathrm{start} = \int_{z_\mathrm{start}}^{z_\mathrm{stop}} \frac{\mathrm{d}z}{v_z(z)},
 \label{equ:tof-integration}
\end{equation}
where $\mathrm{d}z$ is a line element along the $z$ axis and the velocity $v_z$ of the particle is given by
\begin{equation}
v_z(z) =  c\, \sqrt{1 - \dfrac{m_0^2 c^4}{\left(E_\mathrm{kin,\,start}+q(U(z)-U_\mathrm{start}) + m_0 c^2\right)^2}}, \label{equ:velocity-simulation}
\end{equation}
with $q=-e$. For simplicity we assume an electron with solely axial kinetic energy and emitted in the center of the flux tube ($r=0$). For an electron energy just large enough to fulfill the transmission condition given in \ref{equ:tcond} the velocity approaches zero in the analyzing plane where $E_\mathrm{kin}-qU_0 \approx 0$ and its time of flight will approach infinity.

\end{itemize}

The principle of a source that meets the above requirements has already been described in \cite{art_fegun,diss_kathrin}. However, this prototype had only a limited angular selectivity. In the following we will discuss an electron source that fulfills all of the above requirements, including a high degree of angular selectivity. First we will describe the principle of the improved angular-selective source. This section will be followed by measurements that demonstrate the functionality of our new angular-selective electron source, suitable for the in-depth characterization of the transmission function of the KATRIN main spectrometer.

\section{Principle of the angular-selective electron source}

\begin{figure}[b!]
	\centering
		\includegraphics[width=0.48\textwidth]{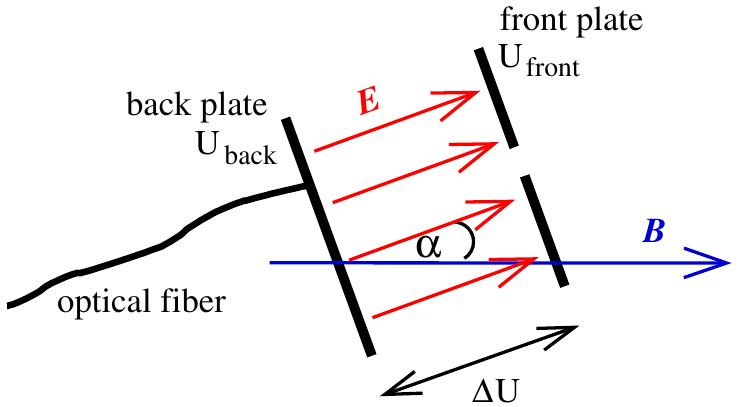}
	\caption[Concept of the new angular-defined electron source]{The schematic concept of the angular-selective electron source proposed in this work (explanation: see text).}\label{fig:plates_concept}
\end{figure}

In \cite{art_fegun} we demonstrated experimentally that an initial transversal energy, \ie\ an angle $\theta_i \ne  \unit[0]{^\circ}$, of the electrons can be achieved using non-parallel electric and magnetic fields. However, the method used in the cited work resulted in a large angular spread of the electrons, which is unsuitable for a detailed investigation of the transmission function with high precision. In order to get electrons with a reasonably small angular spread the geometry close to the emission point has to be flatter than in \cite{art_fegun}. To this end a parallel plate set-up \cite{hugenberg_erice2009,dipl_hein,diss_bokeloh} is used, where the plates are oriented at an angle $\alpha$ with respect to the magnetic field, which is parallel to the spectrometer axis (Fig.~\ref{fig:plates_concept}). The back plate is set to negative high voltage $U_\mathrm{back}$, whereas the front plate is set to a more positive voltage $U_\mathrm{front}$ with a difference voltage of 
\begin{equation}
\Delta U = U_\mathrm{front} - U_\mathrm{back} > \unit[0]{V}
\label{equ:diffv}
\end{equation}
between the two. Electrons are created close to the center of the back plate with the photoelectric effect using UV light \cite{art_fegun,art_uvled}. The UV light is guided through the back plate using a UV transparent optical fiber. The fiber ends at the front surface of the back plate, where it is polished flush with the plate. The front of the fiber is coated with silver for photoelectron production and to ensure that the photoelectrons are all created at the same potential as the back plate. The homogeneous electric field of the parallel plate set-up then assures a well-defined acceleration for all photoelectrons. The photoelectrons have an initial kinetic energy which is given by the energy of the UV photons, the work function of the silver, and the energy distribution of electrons close to the Fermi level of the silver. This is small compared to the energy that the electrons gain in the electric potential. However, the distribution of the initial kinetic energies will contribute to the broadening of the transmission function for high-resolution spectrometers like KATRIN.

Upon being emitted, the photoelectrons instantly experience the Lorentz force $\vec{F}$ in the combined field:

\begin{equation}
\vec{F} = q (\vec{E} + \vec{v} \times \vec{B}) . 
\end{equation}

Initially the electric field dominates due to the tiny start energy of the electrons and hence their low velocity $\vec{v}$. Therefore the electrons will be accelerated along the electric field $\vec{E}$, which is at an angle $\alpha$ with respect to the magnetic field $\vec{B}$. This transfers both axial as well as transversal energy to the electrons. After some acceleration due to the electric field the magnetic part of the Lorentz force will gain influence and the electrons will start to gyrate around the magnetic field lines towards the spectrometer. They leave the parallel plates through an aperture in the front plate.

Due to the homogeneity of the electric field in a parallel plate capacitor the electric field is at the same angle with respect to the magnetic field for all electrons and a well-defined transversal energy is expected for the electrons, \ie\ a narrow interval of gyration angles around the magnetic field.

\subsection{Simulations of the parallel plate set-up}

\begin{figure}
\centering
\subfigure[Increase of the transversal energy between the plates.]{\label{fig:etrans_zoom}
\includegraphics[width=0.7\textwidth,angle=0]{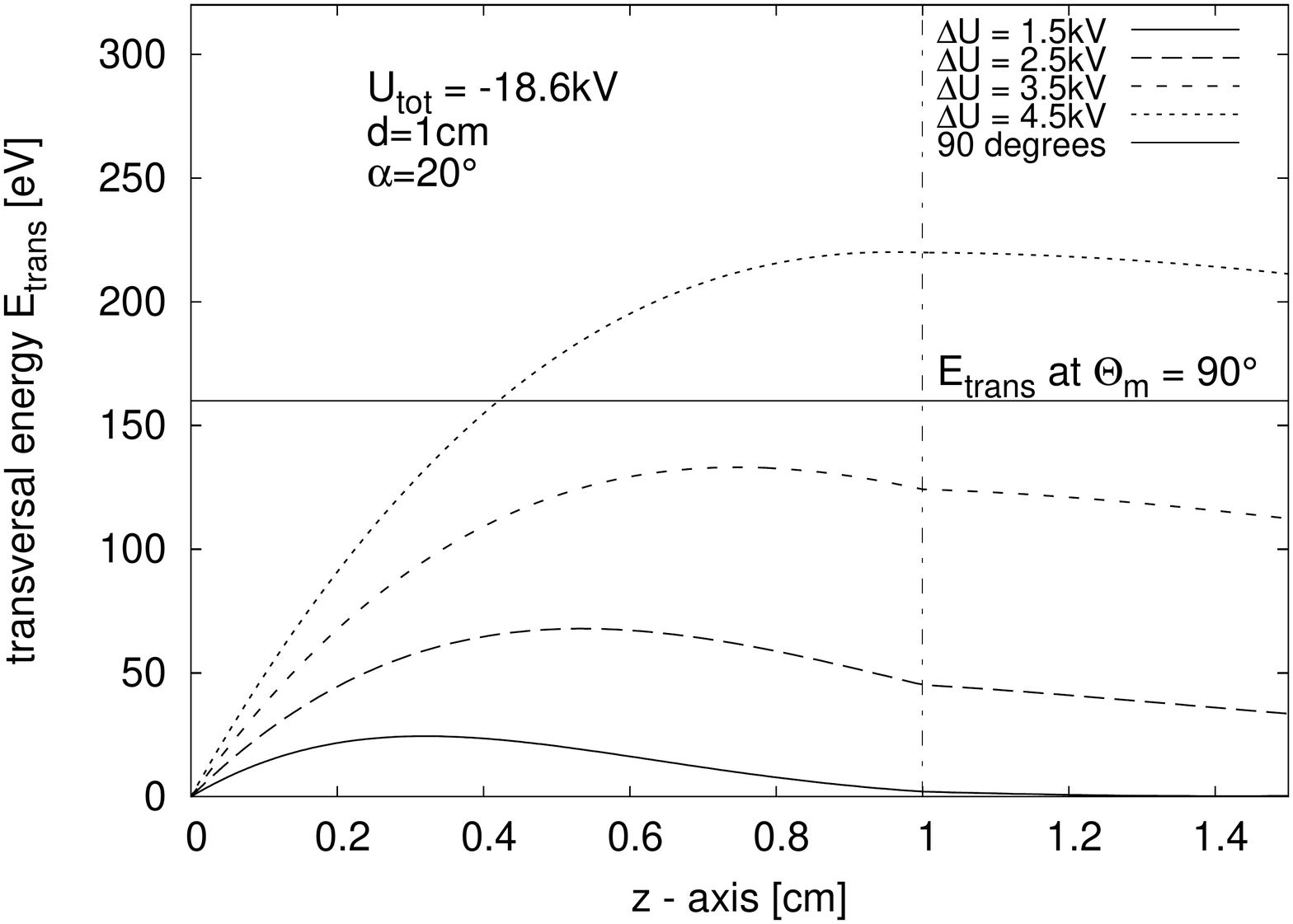}}
\subfigure[Oscillation of the transversal energy outside the parallel plates.]{\label{fig:deltau_etrans}
\includegraphics[width=0.7\textwidth,angle=0]{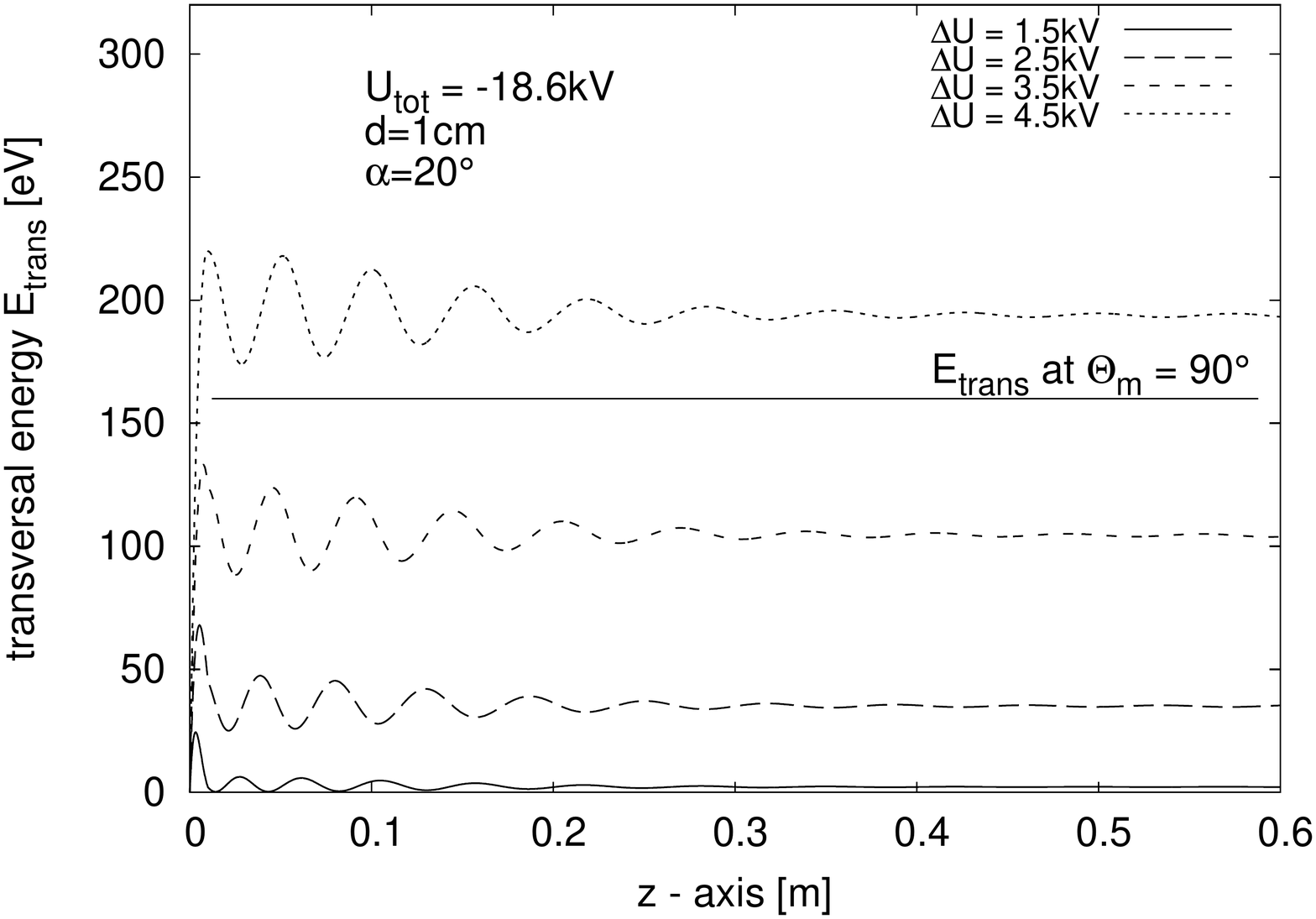}}
\caption[Potential difference dependence]{Numerically calculated evolution of the transversal energy $E_\perp(z)$ of an electron as a function of the accelerating voltage $\Delta U$ (note the different scales of the abscissa in the subfigures). The angle $\alpha=\unit[20]{^\circ}$ is fixed in the simulation. The back plate is positioned at $z_\mathrm{back}=0$ and the front plate at $z_\mathrm{front}=\unit[1]{cm}$, marked by a vertical line. The horizontal line indicates the minimum transversal energy required to obtain electrons of $\theta_m = \unit[90]{^\circ}$ in the entrance solenoid of the spectrometer. (a) Depending on $\Delta U$, the parallel-plate set-up allows to control the transversal energy of the emitted electrons in such a way that angles in the full range from $\theta_m \approx \unit[0]{^\circ}$ to $\theta_m \approx \unit[90]{^\circ}$ are reached at the entrance of the spectrometer. (b) With increasing distance to the plate the oscillations (see text) of $E_\perp$ decrease until, at infinite distance, the transversal energy converges to a constant value $E_{\perp,\infty}$.}
\label{fig:simetrans}
\end{figure}

Particle tracking simulations were used to understand the behavior of the electrons for the set-up described above \cite{dipl_hein,diss_bokeloh}. Figure~\ref{fig:simetrans} shows the simulated development of the transversal energy in and just outside of the parallel plate set-up for typical parameters of the parallel plates used. Initially, the force due to the electric field dominates and leads to a monotonous increase of $E_\perp$ (Fig.~\ref{fig:etrans_zoom}, at small $z$). As the Lorentz force due to the magnetic field gets stronger, the electrons start to gyrate around the magnetic field lines. This periodic motion through the electric field manifests as oscillations of $E_\perp$ in the vicinity of the parallel plate set-up where still a sizable transversal component of the electric field exists (Fig.~\ref{fig:deltau_etrans}, at small $z$). These oscillations vanish far away from the parallel-plate set-up where no sizable transversal electric field is present anymore and the transversal energy converges to a constant value $E_{\perp,\infty}$.

\begin{figure}[h!]
	\centering
        \includegraphics[width=0.7\textwidth]{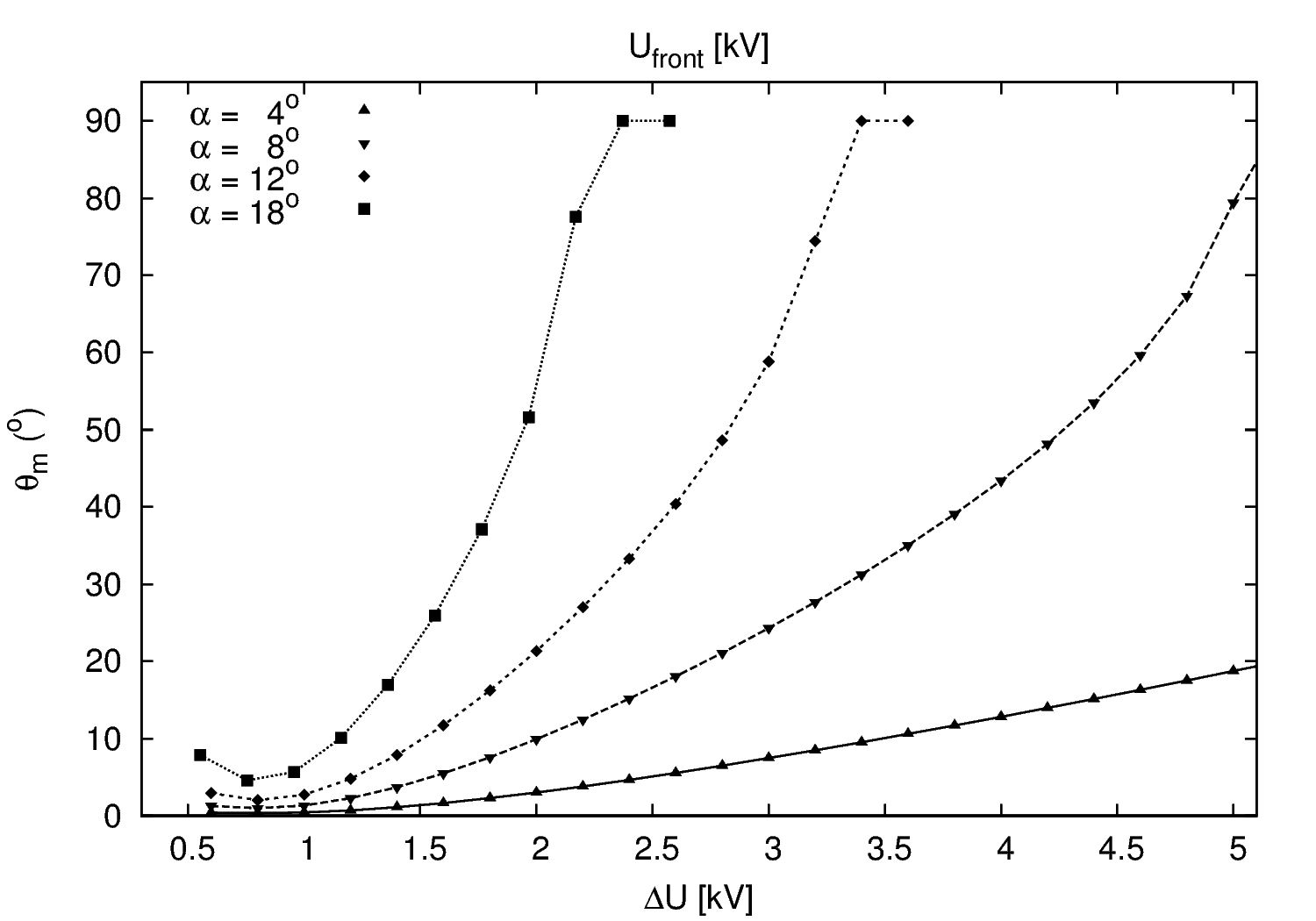}
	\caption[Summary of simulations]{Angle $\theta_m$ in the center of the first spectrometer solenoid for different plate angles $\alpha$ and different 
plate difference voltages $\Delta U=U_\mathrm{front}-U_\mathrm{back}$ from simulations for a fiber at $r=\unit[0.9]{mm}$ off center (explanation see text). The magnetic field at the source was set to $B = 20~\mathrm{mT}$. (Figure adapted from \cite{diss_bokeloh}.)}\label{fig:geoxp-thofu200}
\end{figure}

Transversal energy gains in the region of several hundred eV are achieved. Still, at kinetic energies of $18~\mathrm{keV}$ this corresponds to small angles only. However, similar to the adiabatic conversion of transversal into axial energy in the spectrometer, a small transversal energy can be increased according to Eq.~\ref{equ:angle-transform} by guiding the electrons from a weak magnetic field into a strong magnetic field region. To use this effect the test measurements took place with $B_i \approx 20~\mathrm{mT}$ at the location of the parallel plate set-up and $B_f = B_\mathrm{max} = 6~\mathrm{T}$ at the entrance solenoid of the Mainz MAC-E filter (see section~\ref{sssec:mainz_spec}). This way angles of up to $\theta_m = 90^\circ$ can be achieved at the entrance of the spectrometer. Figure~\ref{fig:geoxp-thofu200} shows the results of simulations for the angle $\theta_m$ for similar experimental conditions as used in the test experiment which is described in the following section. Results are displayed for fixed angle of rotation, $\alpha$, of the plates and variable potential $\Delta U$ between the plates. Clearly, a given angle, \ie \ transversal energy, can be achieved either by varying $\Delta U$ with fixed $\alpha$, or with $\Delta U$ fixed and $\alpha$ varied.

\section{Experimental test of the principle}

\begin{figure*}[htb!]
\centering
\includegraphics[width=1.0\textwidth]{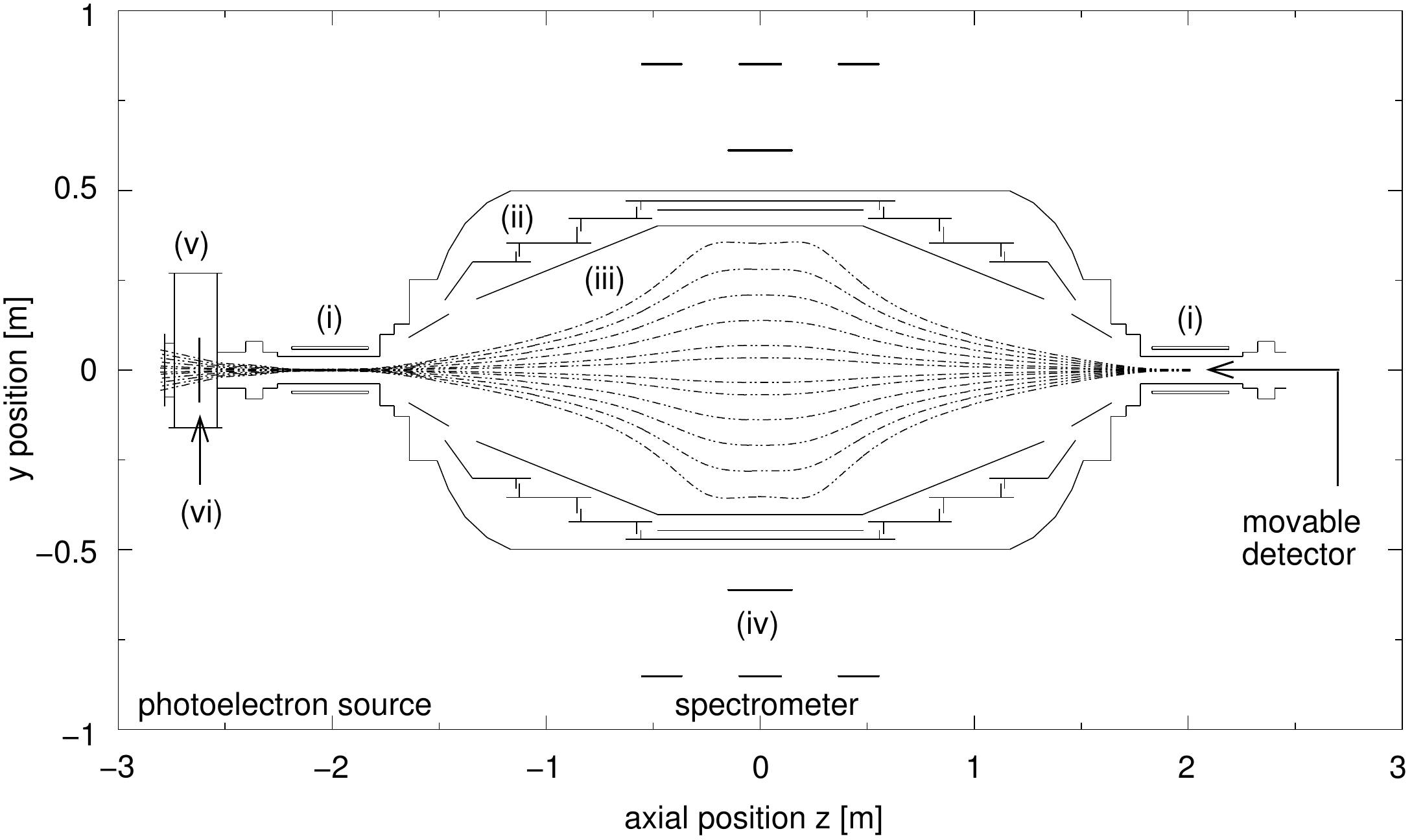}
\caption{Schematic of the experimental set-up to test the photoelectron source. From left to right: vacuum chamber with photoelectron source, MAC-E filter, and detector. The details shown in the sketch include: (i) two superconducting solenoids to produce the magnetic guiding field of the MAC-E filter, (ii) the electrode configuration comprising a vacuum tank on ground potential and an inner high-voltage electrode system, (iii) the innermost wire electrode, (iv) field-shaping air coils, (v) vacuum chamber for the electron source, and (vi) the parallel plates of the photoelectron source. Magnetic field lines connecting the photoelectron source and the detector are indicated as dashed curves. 
(Figure adapted from~\cite{art_uvled}.)} \label{fig:mainz_spectrometer}
\end{figure*}

The principle of the angular-selective electron source was tested at the MAC-E filter of the former Mainz Neutrino Mass Experiment. The system consisted of the parallel-plate set-up at the entrance of the Mainz spectrometer and a detector at its exit (Fig.~\ref{fig:mainz_spectrometer}).

\subsection{Experimental set-up}

\subsubsection{The angular-selective electron source}

\begin{figure}[htbp]
	\centering
		\includegraphics[width=0.48\textwidth]{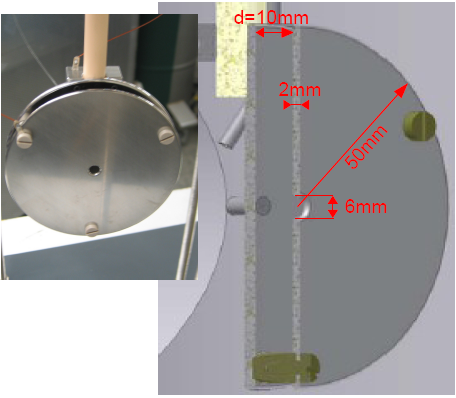}
		\caption{The set-up of the electron source for the measurement at the Mainz spectrometer. The two parallel plates are mounted on a ceramic rod, which is connected to a manipulator (not shown). The manipulator and two high voltage feedthroughs are mounted on three CF 35 flanges. One of the HV feedthroughs is hollow and carries the fiber into the vacuum chamber. It is sealed off with a UHV-compatible epoxy.}
	\label{fig:egun}
\end{figure}

The source with a plate diameter of $\varnothing_\mathrm{plate}=100~\mathrm{mm}$ was set up in a vacuum chamber at the entrance of the Mainz spectrometer. The set-up of the source itself is shown in Fig.~\ref{fig:egun}. The plates are made of stainless steel with a thickness $d_\mathrm{plate}=2~\mathrm{mm}$. The plate distance, which is fixed by three PEEK (polyether ether ketone) screws suitable for the use in ultra high vacuum, is  $d=10~\mathrm{mm}$. The aperture in the front plate has a diameter of $\varnothing_\mathrm{aperture}=6~\mathrm{mm}$.  

The back plate is mounted on a ceramic rod, which insulates the plates from ground potential. A rotational feedthrough is mounted on top of the vacuum chamber and is connected to the ceramic rod. This allows rotation centered around the aperture in the front plate of the source and provides for a precise setting of the angle $\alpha$ with respect to the magnetic field in steps of $\Delta\alpha\approx 0.1^\circ$. 

The optical fiber is mounted through the back plate via a microdrilling at a distance of $0.9~\mathrm{mm}$ from the center, whereas the aperture in the front plate is exactly centered. The tiny offset in radial position of the two allows the photoelectrons to pass through the center of the aperture, taking into account a small deflection of their trajectory due to the magnetic field. This helps to avoid the inhomogeneous electric field close to the edge of the aperture, which would lead to an increased spread of their transversal energy.

A fiber of type FEQ100-15000100 (15/125/250) from j-fiber GmbH, Jena, with a core diameter of $\varnothing_\mathrm{core}=15~\mathrm{\mu m}$ was used. Its end was polished flush with the front surface of the back plate. A silver layer with a density of about $(35 \pm 5)~\mathrm{\mu g / cm^2}$ was evaporated onto the end of the fiber. Silver with a work function of $\Phi_\mathrm{Ag} = (4.6 \pm 0.1)~\mathrm{eV}$ \cite{Ag_workfcn} has proven to be a suitable material for photoemission in conjunction with the UV-LEDs used (wavelength $(265 \pm 15)~\mathrm{nm}$ corresponding to a photon energy of $(4.68 \pm 0.27)~\mathrm{eV}$ \cite{art_fegun,art_uvled,seoul-led-265}). The other end of the fiber was clamped in a metal jacket below a position-adjustable mount for the UV diode.

\subsubsection{The spectrometer}
\label{sssec:mainz_spec}

The Mainz spectrometer (see Fig.~\ref{fig:mainz_spectrometer}) consists of a vacuum tank on ground potential, an inner electrode system to create the analyzing potential, two superconductiong solenoids to create the magnetic field and an air coil system for earth field compensation and  to tune the magnetic field in the analyzing plane. The electrode system comprises a set of several cylindrical retardation electrodes close to the tank walls on negative high voltage, and a grid electrode made of wires \cite{flatt-paper}. In our measurements, the magnetic field inside the superconducting solenoids was set to $B_\mathrm{max} = \unit[6]{T}$. This amounts to a field at the center of the aperture of the electron source of $0.0209~\mathrm{T}$. The field strength in the analyzing plane of the spectrometer was adjusted to $B_\mathrm{min}=1.9~\mathrm{mT}$ using the air coil system. Inserting the values into Eq.~(\ref{equ:energyres}), these settings result in an energy resolution for $15~\mathrm{keV}$ electrons\footnote{The measurements had to be performed at voltages close to $15~\mathrm{kV}$ instead of $18.6~\mathrm{kV}$ since discharges occurred for higher voltages. These were caused by the Penning trap formed by the source, the analyzing potential and the magnetic field (see \cite{art_trap}) and a filling mechanism of this trap caused by a non-optimal mechanical finishing of the front and back plate of the source and their electrical connections. For the measurements in \cite{art_trap} a wire in the beam line a bit off axis was used to remove stored electrons.} of $\Delta E  = 4.7~\mathrm{eV}$. This resolution was chosen significantly broader than the $0.9~\mathrm{eV}$ that can be achieved with the Mainz spectrometer (see \cite{art_uvled}) in order to enhance the effect of different transversal energies on the transmission functions (see Eq.~\ref{equ:energyres}), \ie\ to maximize the interesting effect and to minimize disturbances.

Since the energy selection is performed by the electrostatic filter, the
electron detector merely serves as a counter to measure an integrated spectrum
of those electrons that pass the retardation potential at a specific potential setting $U_0$. We chose a windowless Si-PIN diode (type Hamamatsu S3590-06)
of size $\unit[9 \times 9]{mm}^2$ as electron detector.

\subsubsection{High voltage set-up}

Three voltages are needed to measure a transmission function with the angular-selective electron source. The main voltage of $U_\mathrm{back}\approx -15~\mathrm{kV}$ was directly connected to the back plate of the angular-selective electron source. It defines the kinetic energy of the photoelectrons. The front plate of the electron source was connected to the back plate via a potential difference $\Delta U$ in the range $[0~\mathrm{V},+5~\mathrm{kV}]$, resulting in $U_\mathrm{front} = U_\mathrm{back} + \Delta U$. $\Delta U$ is one of the parameters which determine the angle $\theta_m$ of the photoelectrons (see Fig.~\ref{fig:geoxp-thofu200}). The retarding electrode system of the spectrometer was connected to the back plate via a variable voltage $U_\mathrm{scan}$, resulting  in $U_0 = U_\mathrm{back} + U_\mathrm{scan}$. This scan voltage ranged from slightly negative values of $\mathcal{O}(-10~\mathrm{V})$ to $\approx +500~\mathrm{V}$ to scan the transmission function from full reflection ($U_\mathrm{scan}<0$) to transmission at high excess energy ($U_\mathrm{scan}>0$).

\subsection{Measurement of transmission functions for different transversal energies}

\begin{figure}[ht!]
	\centering
		\includegraphics[width=.7\textwidth]{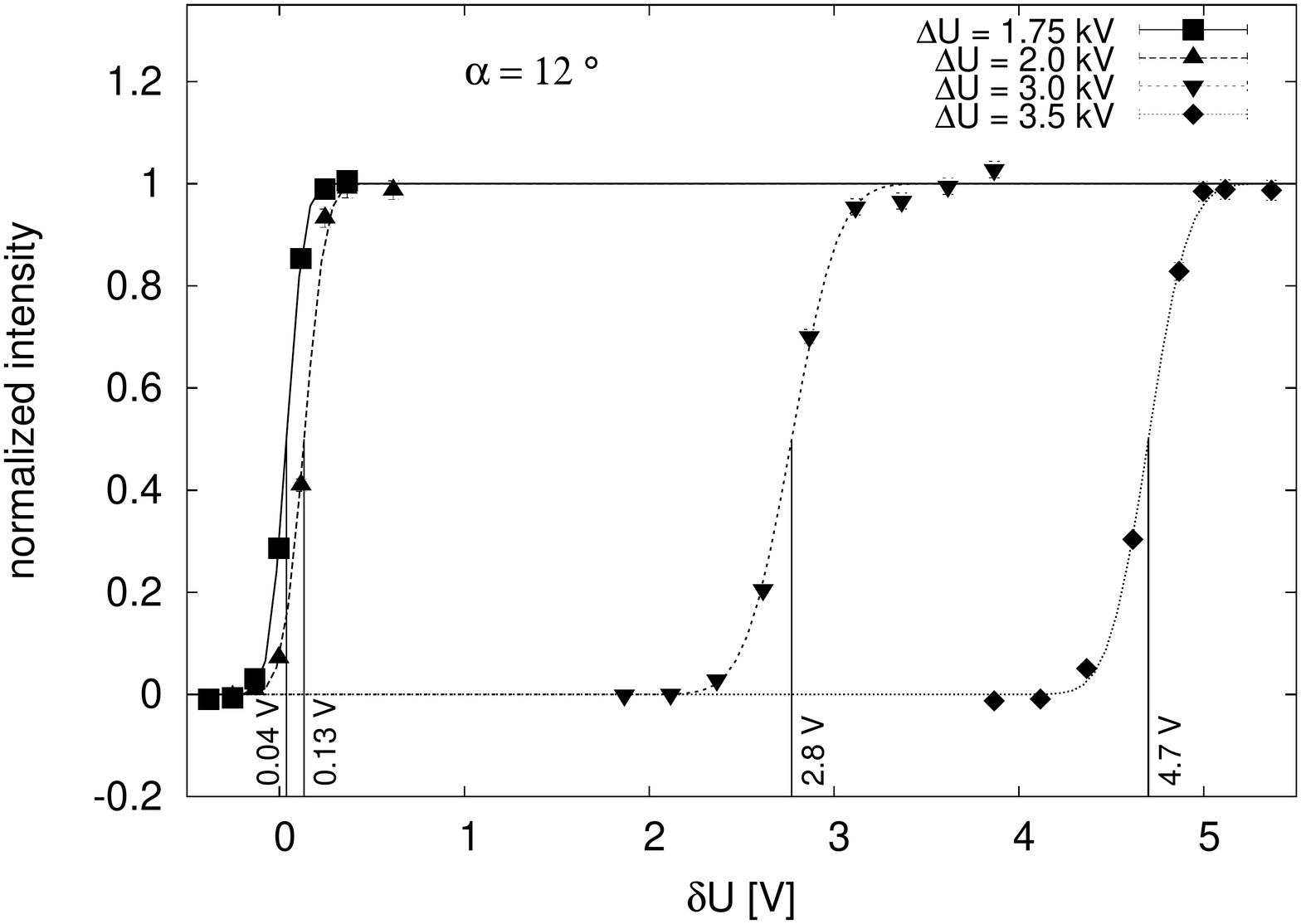}
	\caption{Transmission functions for $\alpha=\unit[12]{^\circ}$ and four different $\Delta U$.  The functions were fitted with an error function. The centers of the functions resulting from the fit are illustrated as vertical labels. Below $\Delta U=\unit[1.75]{kV}$ and above $\Delta U=\unit[3.5]{kV}$ no photoelectrons were detected.}
	\label{fig:plot_12deg_all}
\end{figure}

\begin{figure}[ht!]
	\centering
		\includegraphics[width=.7\textwidth]{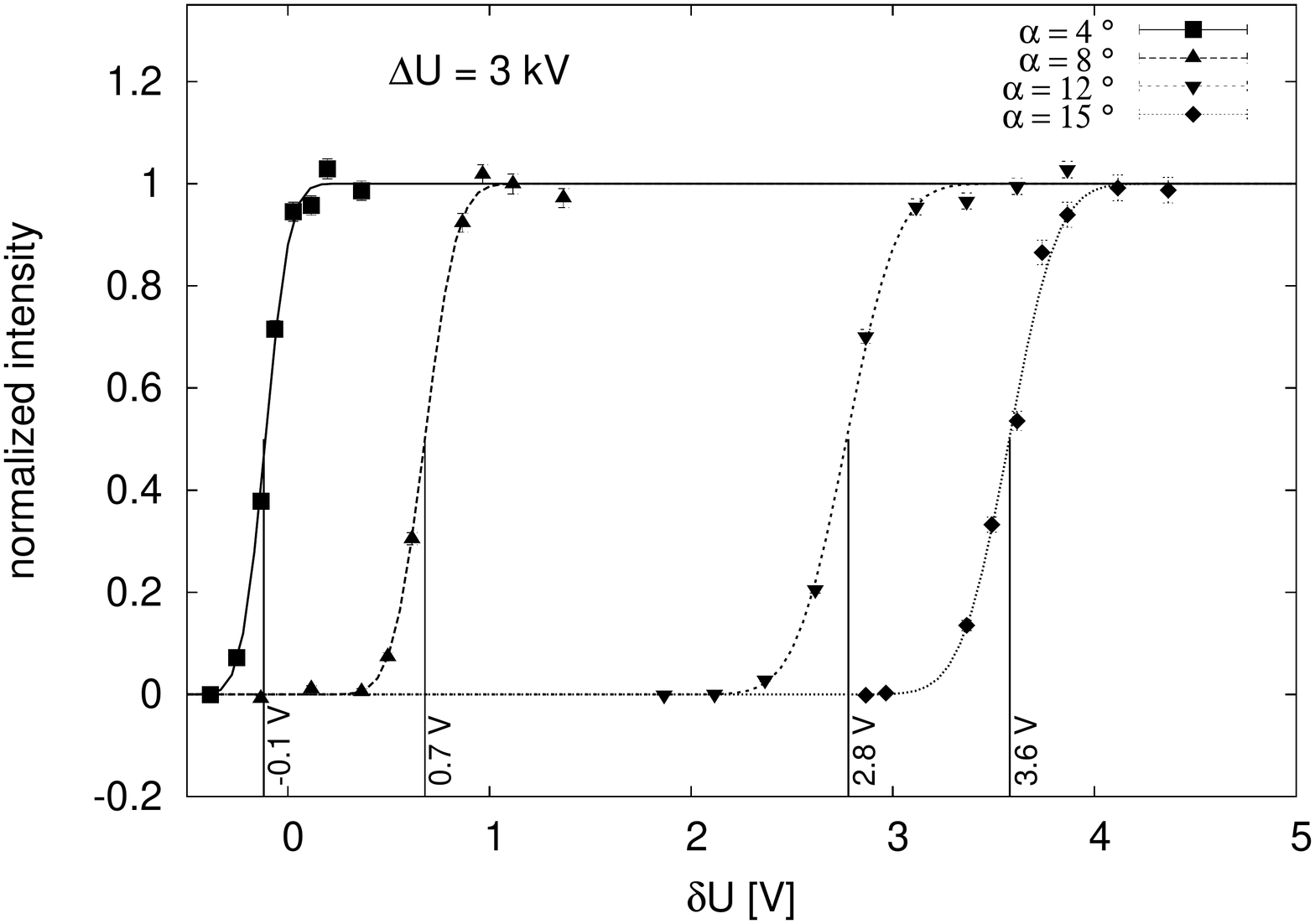}
	\caption{Four transmission functions, measured at $\Delta U=3~\mathrm{kV}$ and different angles $\alpha$. As in Fig.~12 their centers from a fit with an error function are illustrated. Since the measurement was stopped at $\alpha=\unit[15]{^\circ}$ the maximal $\delta U$ was not reached.}
	\label{fig:plot_3kV_all}
\end{figure}

We measured transmission functions with the above set-up for various voltage differences $\Delta U$ and angles of rotation $\alpha$ (Figs.~\ref{fig:plot_12deg_all} and \ref{fig:plot_3kV_all}). For ease of comparison of these transmission functions with respect to the resolution of the spectrometer we use voltage differences $\delta U$ to characterize their position with respect to the transmission threshold for which only electrons with purely axial energy are transmitted: $\delta U = 0~\mathrm{V}$ is the transmission threshold and $\delta U = \Delta E/e = 4.7~\mathrm{V}$ when all electrons with angles up to $\theta_m = 90^\circ$ are being transmitted. It follows that $e \delta U$ is the transversal energy of the electrons in the analyzing plane.  The transmission function for electrons with angles close to $\theta_m = 90^\circ$, \ie, $E_\perp = 4.7~\mathrm{eV}$ in the analyzing plane and therefore $\delta U = 4.7~\mathrm{V}$, was experimentally determined by setting the spectrometer to full transmission for all angles and ramping up $\Delta U$ until barely any electrons were counted anymore. The corresponding transmission function in Fig.~\ref{fig:plot_12deg_all} is the rightmost curve and defines the $x$-axis. It is achieved for $\Delta U = 3.5~\mathrm{kV}$ for $\alpha = 12^\circ$. This is perfectly consistent with the simulation (Fig.~\ref{fig:geoxp-thofu200}) for which, at $\alpha = 12^\circ$, $\theta_m = 90^\circ$ is reached at $\Delta U =3.5~\mathrm{kV}$\footnote{The difference of the back plate potential ($15~\mathrm{kV}$ for the measurements instead of $18~\mathrm{kV}$ for the simulation) is of no consequence since the radial energy does not depend on the absolute electric potential but solely on $\Delta U$ and $\alpha$.}.

The measured transmission functions show widths that are much less than the resolution of the spectrometer of $\Delta E = 4.7~\mathrm{eV}$ and their centers cover a range of $4.8~\mathrm{eV}$ which, allowing for some experimental uncertainties, is the same as the resolution (Figs.~\ref{fig:plot_12deg_all}, \ref{fig:plot_3kV_all} and Table~\ref{tab:summary_plates}). This clearly shows that electrons with different and well-defined transversal energies, \ie\ start angles, have been created at the different settings of plate rotation angle $\alpha$ and potential difference between the plates $\Delta U$.

\begin{table}[ht!]
	\centering
	\caption{Position $\delta U$ and width $\sigma_\mathrm{TM}$ of the measured transmission functions. The fit was performed using an error function. The uncertainty of $\Delta U$ was $0.1~\mathrm{V}$ and of $\sigma_\mathrm{TM}$ $0.01 -0.02~\mathrm{V}$. The uncertainties given result from the fit only.}

	\label{tab:summary_plates}
		\begin{tabular}{|c|c|c|c|}
			\hline
			$\alpha$ ($\pm\unit[0.1]{^\circ}$) & $\Delta U$ [kV] & $\delta U$ [V] & $\sigma_\mathrm{TM}$ [V] \\
			\hline
			\hline
			$\unit[12]{^\circ}$ &1.75 & 0.04 (0.003) & 0.076 \\
			$\unit[12]{^\circ}$ &2.0 & 0.13 (0.009) & 0.10 \\
			$\unit[12]{^\circ}$ &3.0 & 2.8 (0.011) & 0.2 \\
			$\unit[12]{^\circ}$ &3.5 & 4.7 (0.011) & 0.17 \\
			\hline
			$\unit[4]{^\circ}$ & 3.0 & -0.1 (0.005) & 0.10 \\
			$\unit[8]{^\circ}$ & 3.0 & 0.7 (0.009) & 0.13 \\
			$\unit[12]{^\circ}$ & 3.0 & 2.8 (0.011) & 0.20 \\
			$\unit[15]{^\circ}$ & 3.0 & 3.6 (0.011) & - \\
			\hline
		\end{tabular}
\end{table}

\subsubsection{Position of the measured transmission functions}

Figures~\ref{fig:plot_12deg_all} and \ref{fig:plot_3kV_all} show the measured transmission functions. Electrons with small emission angles, \ie\ emission with most of the energy in the axial component, are measured at small $\delta U$. In contrast, electrons emitted at large angles with most of their energy in the transversal component are measured at large $\delta U$. The positions of the measured transmission functions increase monotonically with $\Delta U$ and $\alpha$, as is expected from the simulations. The spread of the positions of the measured transmission functions between $e\delta U \approx 0~\mathrm{eV}$ and $e\delta U = 4.7~\mathrm{eV}$ shows that the angle $\theta_m$ at the entrance of the spectrometer can be created in the interval of $0^\circ$ to $90^\circ$. These observations confirm that the transversal energy of the photoelectrons can be selected by the parameters $\alpha$ and $\Delta U$.

\begin{figure}[hbtp]
	\centering
		\includegraphics[width=0.7\textwidth,angle=0]{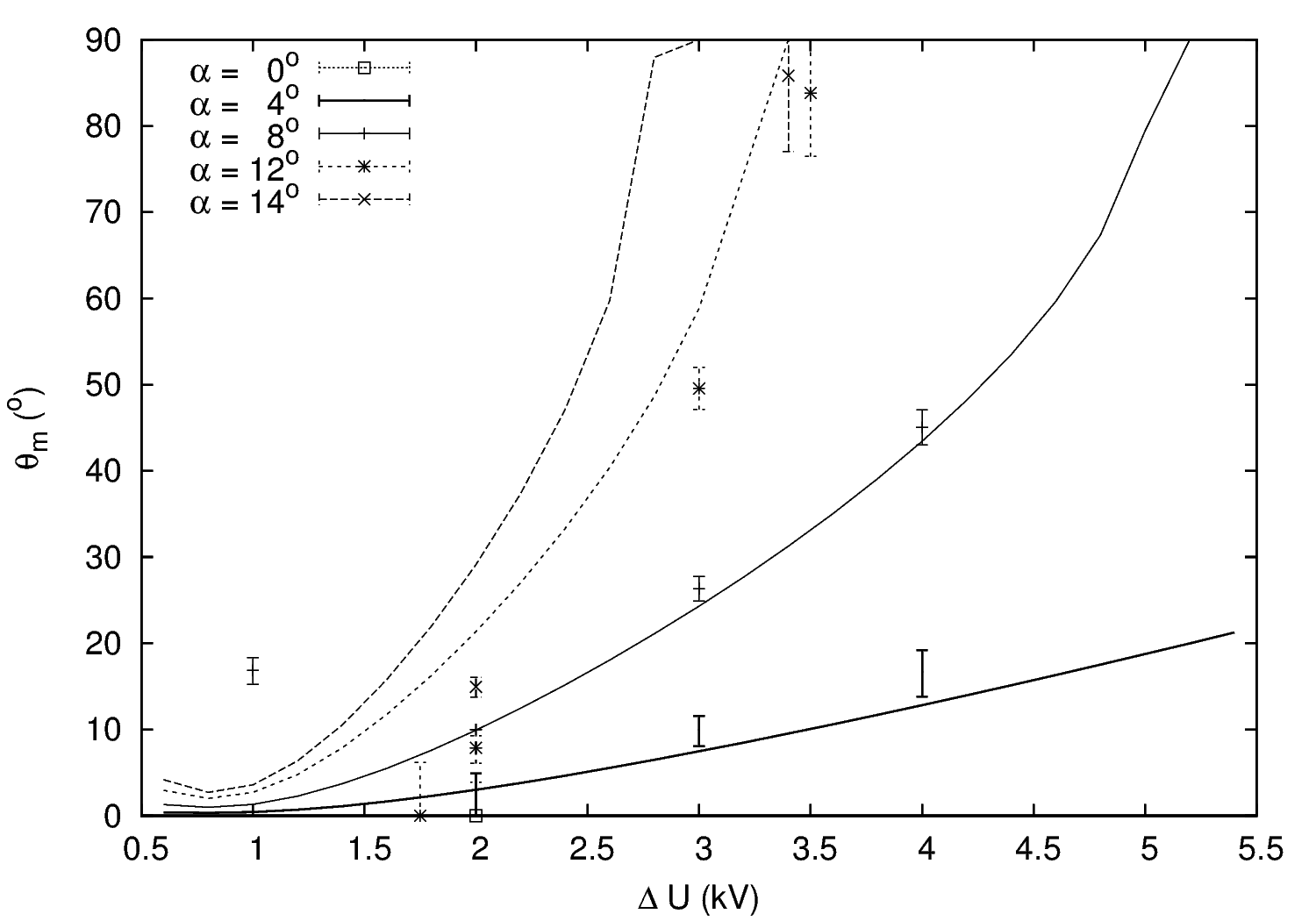}
	\caption{For five angles $\alpha$ measurements and simulations of $E_\perp = e \delta U$ in the analyzing plane are compared (lines for the simulations, symbols for the measurement). In general, the simulations describe the measurements best at large $\Delta U$ and small angles $\alpha$. (Figure adapted from \cite{diss_bokeloh}.)}
	\label{fig:plot_sim_and_mes_all}
\end{figure}

In order to understand the measured transmission functions they are compared with simulations in Fig.~\ref{fig:plot_sim_and_mes_all}. Here both simulations (lines) and measurements (symbols) are shown together. The simulations follow the measurements qualitatively, but fail to describe them in detail. This is not unexpected since the measurements were performed in a non-cylindrical vacuum chamber whereas the simulations assume a cylindrically symmetric geometry due to constraints in the simulation software. Deviations from cylinder symmetry will lead to transversal components of the E-field which still may alter $E_\perp$ slightly. Additionally, the position of the source in the magnetic field was not known with high precision, causing a potential deviation of the actual magnetic field strength and degree of homogeneity compared to the values used in the simulation.

\subsubsection{Width of the measured transmission functions}

In addition to their position the measured transmission functions are characterized by their width. Ideally, the transmission function for electrons of a single, defined angle is a step function (Fig.~\ref{fig:tfuncsigma}). However, several effects contribute to an additional broadening. The dominant of these are a finite angle interval due to inhomogeneous E- and B-fields at the source, a finite distribution of starting energies, and electronic noise on the scanning voltage applied between the back plate and the spectrometer. The broadening of the angular distribution due to the inhomogeneity of the E- and B-fields is determined by the geometry of the set-up and as such inherent in the set-up used. A finite distribution of starting energies is caused by the spectral width of the UV light and a spread of the work function of the back plate across the emission spot, each estimated to lead to a Gaussian broadening with sigma of the order of $0.2~\mathrm{eV}$, see ref.~\cite{art_uvled}. However, the two components can not simply be added, since they are not independent. If the energy spread of the UV light overlaps with the range of the work function the resulting broadening is less than the square of the individual components, since part of the light does not lead to photoionization. The spread of the work function itself consists of several components. In the present set-up a minimum, irreducible component is the thermal broadening of the electron occupancies at the Fermi level at room temperature of $\mathrm{FWHM} \approx 0.2~\mathrm{eV}$.

\begin{figure}[htb]
	\centering
		\includegraphics[width=0.7\textwidth]{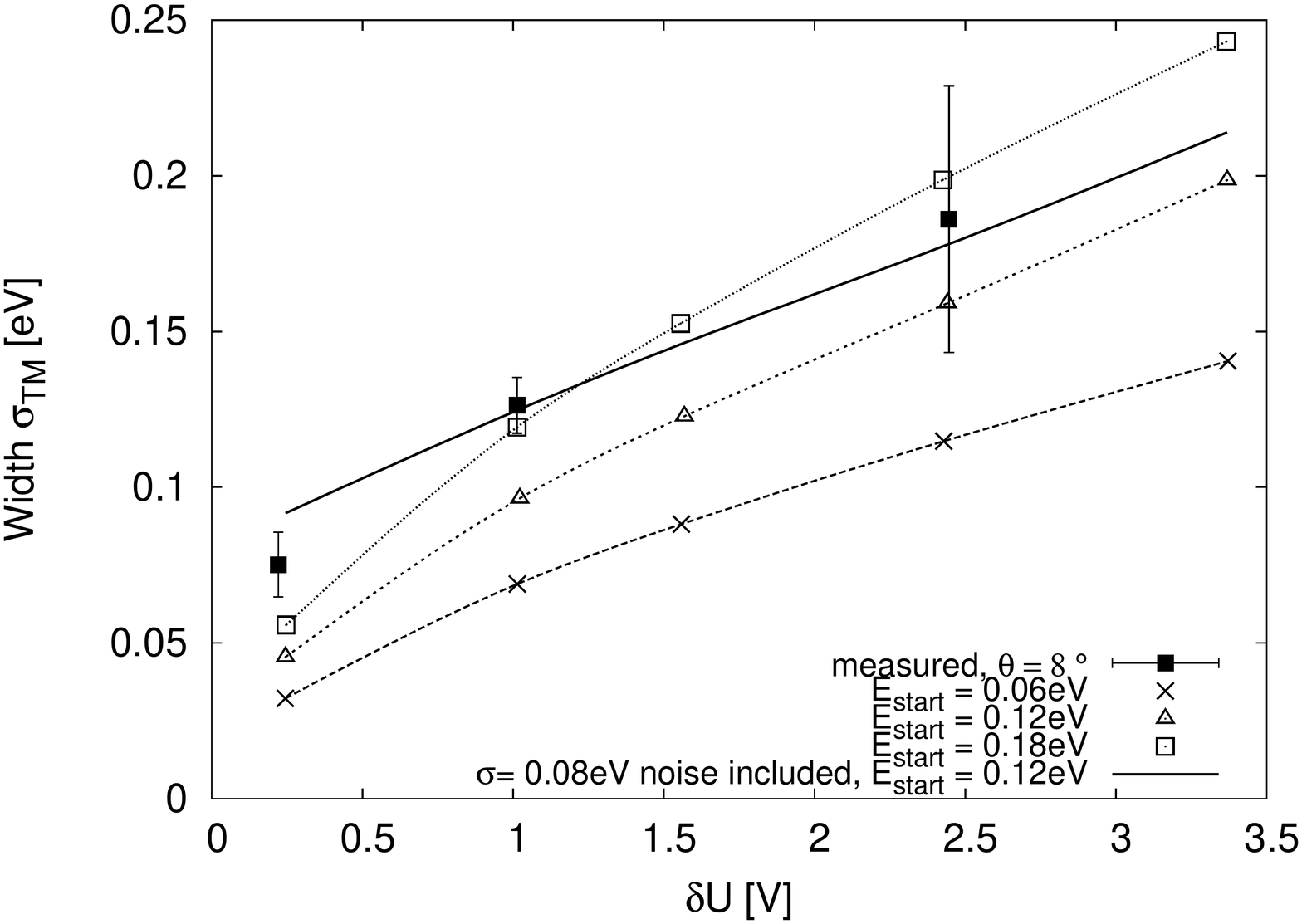}
	\caption{The measured width of the transmission function $\sigma_{\mathrm{TM}}$ as a function of the relative voltage $\delta U$ is compared with simulations for four different starting kinetic energies $E_\mathrm{start}$ of the photoelectrons. The simulated points are spline fitted. For Gaussian noise with $\sigma_\mathrm{noise}=\unit[80]{meV}$ and a starting energy of $E_\mathrm{start}=\unit[0.12]{eV}$ the simulation (solid line) can describe the measured data (filled squares). Both a finite start energy and noise are needed to describe the data.}
	\label{fig:plot_sim_and_mes_xdeg}
\end{figure}

 The consequences of both the inhomogeneous fields and the finite distribution of starting energies on the widths of the measured transmission functions can be quantified using tracking simulations. Several different starting energies of the photoelectrons were used. The direction of emission of the photoelectrons from the surface was assumed to be isotropic. Noise on the applied voltages was included in the simulations using a Gaussian distribution, the width of which was adjusted to yield a good fit with the measurements. This is shown in Fig.~\ref{fig:plot_sim_and_mes_xdeg} for ($\alpha=8^\circ$, $\Delta U = 3.5~\mathrm{kV}$). The simulations can describe the data for an average starting energy of $E_\mathrm{start}=0.12~\mathrm{eV}$ together with a noise on the high voltage at a level of $\sigma=0.08~\mathrm{eV}$, consistent for all three angles $\alpha$ measured (not shown). 

The average starting energy of $E_\mathrm{start}=0.12~\mathrm{eV}$ deviates from the difference between the energy of the UV photons $E_{\text{UV-LED}} = \unit[(4.68 \pm 0.27)]{eV}$, where the uncertainty specifies the spectral width, and the average work function of silver at $\Phi_{\mathrm{Ag}} = 4.6~\mathrm{eV}$. More detailed investigations are ongoing.

This analysis also shows which steps have to be taken to reduce the width of the transmission function, \ie\ the angular distribution of the photoelectrons, even further. Since all three components, spectral spread, work function spread and noise, are at the level of $0.1~\mathrm{eV}$, all have to be reduced in order to achieve a significantly smaller width of the transmission function. The spectral width can be reduced by a better matching of the spectrum of the UV light and the work function distribution, or by using a laser. To improve the spread of the work function the thermal broadening of the electron occupancies at the Fermi level has to be reduced. This is possible in a simple way by cooling the emission spot. The noise on the high voltage at KATRIN will be significantly lower than the numbers above since at KATRIN the high voltage will be stabilized at the level of $< 3~\mathrm{ppm}$ long-term (months), \ie~$< 0.05~\mathrm{V}$, and the short-term fluctuations will be less than $ 0.05~\mathrm{V}$ as well, in contrast to the high voltage set-up used for the measurements presented here.

\subsubsection{Time-of-flight distributions}

Besides the measurements of the transmission function in the energy domain also some time-of-flight distributions were measured. A detailed discussion of the time-of-flight spectra in a MAC-E filter is beyond the scope of this article. However, in order to show that the angular-selective electron source discussed above fulfills all requirements laid down in section~\ref{sec:intro} and in \cite{art_fegun} some time-of-flight measurements will be discussed in the following. A more detailed investigation of time-of-flight distributions using an improved parallel plate electron source will be performed using a dedicated experiment at the main spectrometer of KATRIN.

\begin{figure}[htbp]
	\centering
		\includegraphics[width=0.7\textwidth,angle=0]{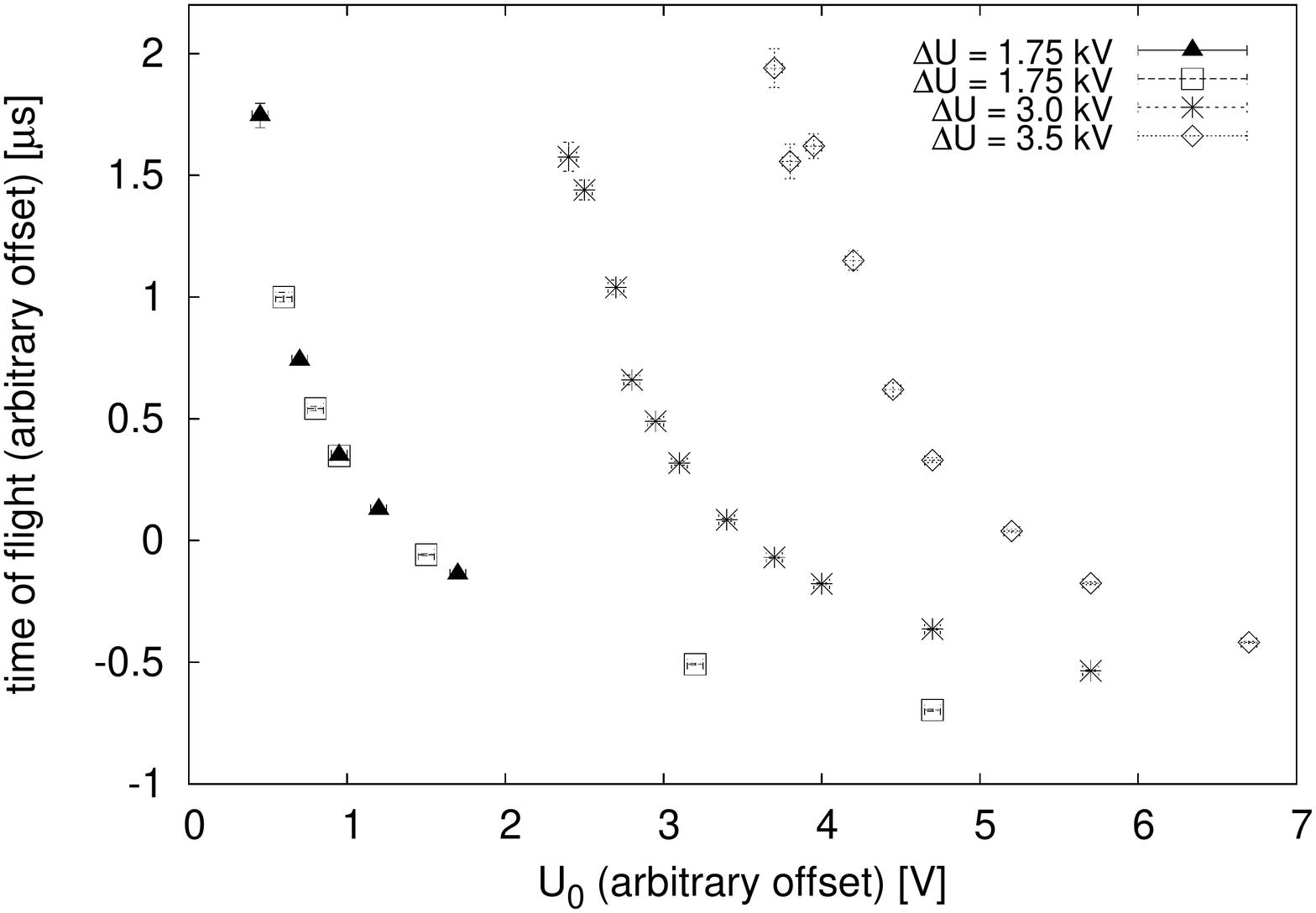}
\caption{Time-of-flight spectra for three of the settings of the corresponding transmission function measurements at $\alpha = 12^\circ$ and $U_\mathrm{back}=-15~\mathrm{kV}$.}
	\label{fig:tof12Grad}
\end{figure}

The data acquisition system for the time-of-flight measurement was based on a Flash-ADC and similar to the one used in \cite{art_uvled}\footnote{A $200~\mathrm{MHz}$ flash ADC was used instead of a $100~\mathrm{MHz}$ flash ADC.}. The UV-LED was used in pulsed mode with pulse lengths of $200~\mathrm{ns}$ and a repetition rate of $50~\mathrm{kHz}$. The start time for the time-of-flight measurements was derived from the trigger of the UV-LED pulse and the stop time from the detector signal. Different start angles of the photoelectrons lead to different time of flight values through the spectrometer (Eq.~\ref{equ:tof-integration}). Figure~\ref{fig:tof12Grad} shows the time-of-flight spectra for three different start angles defined by the plate rotation angle $\alpha = 12^\circ$ and three potential differences $\Delta U$. The corresponding transmission functions for these were already shown in Fig.~\ref{fig:plot_12deg_all}. The three time-of-flight spectra exhibit the same shape but start at different voltages. These time-of-flight measurements demonstrate the usability of this angular-selective electron source for time-of-flight studies.

\section{Conclusion}

We have developed an angular-selective electron source suitable for detailed investigations of the transmission function of a high-resolution electron spectrometer. It uses photoelectrons that are created from a thin silver layer at the end of an optical fiber at the back plate of a parallel plate capacitor by UV photons. Due to a tilt of the electric field of the parallel plate capacitor with respect to the guiding magnetic field the photoelectrons gain a defined transversal energy such that the spectrometer can be investigated with electrons of defined angle with respect to the magnetic field.

This electron source fulfills the requirements laid down in \cite{art_fegun} for the investigation of the transmission properties of the main spectrometer of the KATRIN experiment. Especially, (1) the total energy of the electrons is determined by the potential applied to the back plate of the source and can therefore easily be controlled. The energy spread of $\sigma \approx 0.1~\mathrm{eV}$ is lower than the resolution of the main spectrometer of KATRIN $\Delta E_{\mathrm{KATRIN}} = 0.93~\mathrm{eV}$ and can still be improved. (2) Single electrons can be produced, see \cite{art_uvled}. (3) The size of the emission spot is of order $0.1~\mathrm{mm}$, \ie\ effectively point-like. This has been achieved by using optical fibers to guide the UV light for photoelectron production. (4) The source can be mounted on an $x$-$y$ manipulator and is therefore capable of scanning a large area. This facilitates the investigation of the transmission properties across the full flux tube leading through the spectrometer to the detector. For KATRIN this provides the possibility to test the transmission properties from on-axis up to close to the wire electrodes. Consequently, field inhomogeneities in the vicinity of these wire electrodes can be investigated. (5) The angle $\theta_m$ of the electrons at the entrance of the spectrometer can be adjusted between $0^\circ$ and $90^\circ$ by variation of either the voltage difference $\Delta U$ between the two plates of the parallel plate capacitor or its tilt angle $\alpha$. (6) Finally, pulsed operation of the electron source is possible with pulse lengths down to $40~\mathrm{ns}$ \cite{art_uvled}, opening up the possibility of time-of-flight studies of the transmission properties of the spectrometer, which yield complementary information to the transmission functions in the energy domain.

Overall this electron source is suitable for the investigation of the main spectrometer of KATRIN, with emphasis on deviations from the ideal configuration and on systematic uncertainties. Improved versions of the set-up of the electron source described above are under investigation for use at KATRIN. It is planned to use them for the characterization of the transmission properties of the main spectrometer during its comissioning as well as in the rear section at the upstream end of KATRIN to test the transmission through the full experimental set-up during data taking with tritium.

Notably, many of the questions discussed here for the KATRIN experiment arise in a similar way also at other spectrometers. The angular-selective electron source discussed above does not utilize any properties specific to KATRIN, except an initial gradient of the magnetic field. Therefore an application of this angular-selective electron source is also conceivable at other experiments.

\section{Acknowledgements}

This work was supported by the German Federal Ministry of Education
and Research under grant number 05A08PM1. One of the authors (M.\,Z.) was also supported by the Czech Ministry of Education under the contract LA318. We wish to thank the members of AG Quantum/Institut f\"ur Physik, Johannes Guten\-berg-Universit\"at Mainz, for their kind hospitality and for giving us the opportunity to carry out these measurements in their laboratory.

\end{document}